\documentclass[12pt,preprint]{aastex}

\newcommand\um{\ifmmode{\mu{\rm m}}\else{$\mu$m}\fi}
\begin{document}
\title{Deep Mid-Infrared Silicate
Absorption as a Diagnostic of Obscuring Geometry Toward Galactic Nuclei}

\author{N. A. Levenson\altaffilmark{1}, 
M. M. Sirocky\altaffilmark{1}, 
L. Hao\altaffilmark{2},
H. W. W. Spoon\altaffilmark{2}, 
J. A. Marshall\altaffilmark{2}, 
M. Elitzur\altaffilmark{1}, 
and J. R. Houck\altaffilmark{2}}

\altaffiltext{1}{Department of Physics and Astronomy, University of Kentucky,
Lexington, KY 40506; levenson@pa.uky.edu}
\altaffiltext{2}{Astronomy Department, Cornell University, Ithaca, NY 14853-6801}

\begin{abstract}
The  silicate cross section peak near 10\um{}
produces emission and absorption features in
the spectra of dusty galactic nuclei observed
with  the \textit{Spitzer Space Telescope}.
Especially in ultraluminous infrared galaxies, the 
observed absorption feature can be extremely deep,
as IRAS 08572+3915 illustrates.
A foreground screen of obscuration cannot
reproduce this observed feature, even at
large optical depth. 
Instead, the deep absorption requires 
a nuclear source to be deeply embedded in 
a smooth distribution of material that is both
geometrically and optically thick.
In contrast, a clumpy medium can produce only shallow absorption
or emission, which are characteristic of optically-identified
active galactic nuclei.
In general, the geometry of the dusty region and the total
optical depth,
rather than the grain composition or heating spectrum,
determine the silicate feature's observable properties. 
The apparent optical depth calculated from the
ratio of line to continuum emission
generally fails to accurately measure the true optical depth.
The obscuring geometry, not the nature of the embedded source, also
determines the far-IR spectral shape.
\end{abstract}
\keywords{galaxies: active --- galaxies: nuclei --- infrared: galaxies ---
radiative transfer}

\section{Introduction\label{sec:intro}}
The observable spectra of the nuclei of galaxies 
depend on  both the underlying emission sources
and the subsequent obscuration and reprocessing of
their light by material along the line of sight.  
Dust is common in these nuclear regions,
so its spectral signature is often observed
at mid-infrared (mid-IR) wavelengths.  In particular,
the silicate cross section peak  near 10\um{}
produces a spectral feature that 
is observed frequently  in absorption \citep{Roc91,Spo02,Spo04,Wee05} and
sometimes in emission \citep{Hao05,Sie05,Stu05}.
\citet{Hao06} examine local active galactic nuclei (AGNs)
and ultraluminous infrared galaxies (ULIRGs)
with the striking result that 
the average spectra of various classes of galaxies 
show marked differences in 
the 10\um{}  silicate feature (their Fig. 1).
We seek a physical explanation for the differences among these
families.  Moreover, the underlying origin
of the distinct classes must also account for 
the most extreme example, 
the deep absorption of  the ULIRG IRAS 08572+3915.

\section{Observations with the Spitzer Space Telescope\label{sec:obsv}}

\begin{figure}
\centerline{\includegraphics[height=7in]{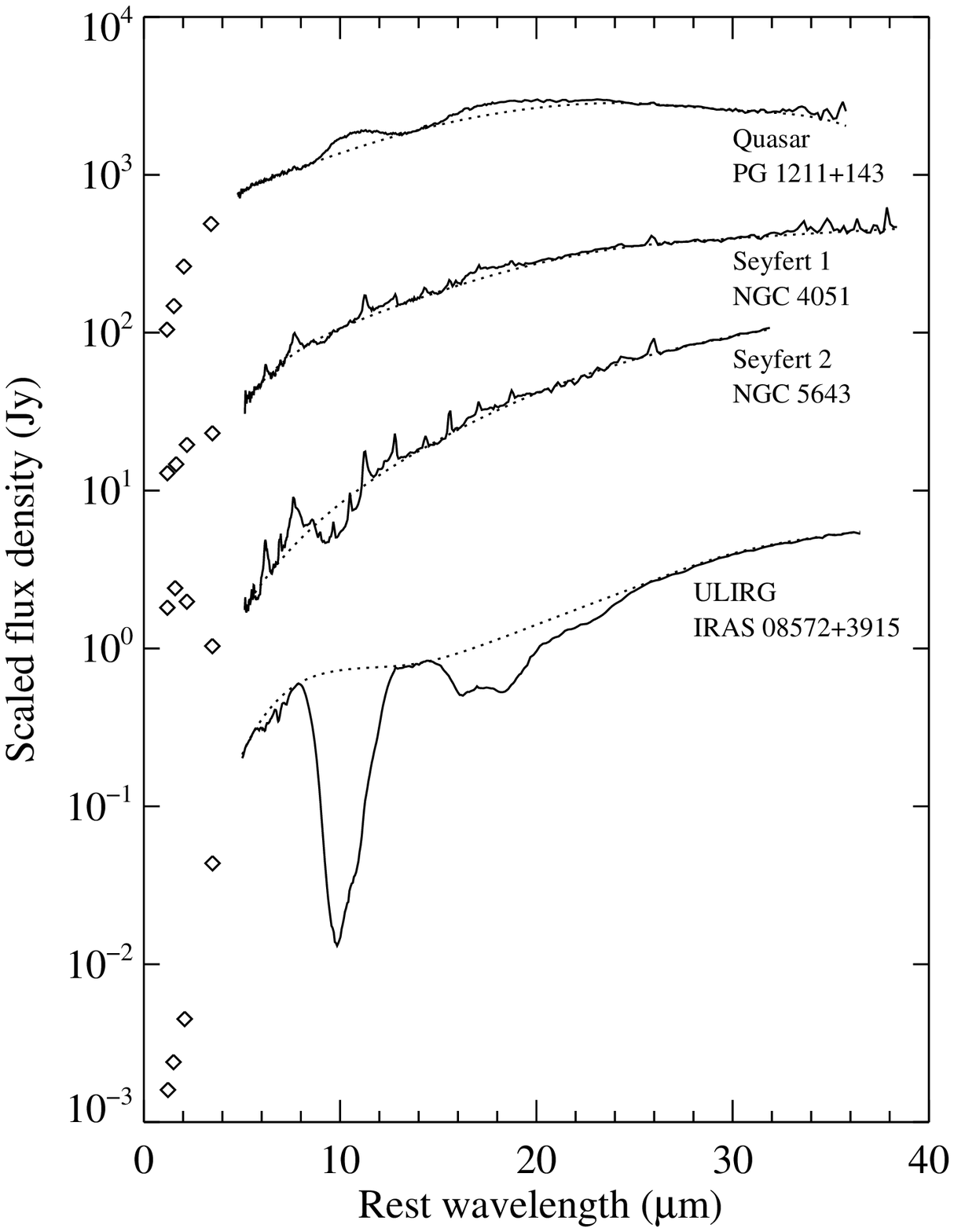}}
\caption{\label{fig:iras08572}
\textit{Spitzer} spectra of the four galaxy classes  
(\textit{solid lines}) and their fitted continua (\textit{dotted lines}). 
The quasar, Seyfert 1, and Seyfert 2 examples are 
typical of their classes.
These plotted
spectra are scaled by factors of $10^4$, 300, and 30, respectively. 
The spectrum of
IRAS 08572+3915 is not scaled, and this galaxy 
 exhibits the most extreme absorption,  
with $S_{sil} = -4.0$.
These mid-IR spectra have been published by 
Shi et al. (2006; PG 1211+143 and NGC 5643)  
and  Spoon et al. (2006a; IRAS 08572+3915). 
Near-IR data (\textit{diamonds}) are from 
\citet{Neu87},
\citet{Rie78},
\citet{Gla85}, and
\citet{Car88}.
}
\end{figure}

\citet{Hao06} 
present a local sample 
of AGNs and ULIRGs observed with the low-resolution modules of
the  Infrared Spectrograph \citep{Hou04} aboard the 
\textit{Spitzer Space Telescope} \citep{Wer04}.
They 
divide the sample into four classes: ULIRGs based on IR luminosity,
and 
quasars, Seyfert 1 galaxies, and Seyfert
2 galaxies based on optical criteria. 
Although some galaxies belong to more than one group,
the average spectra of the groups are very different. 
The silicate feature emerges clearly in emission in 
the quasar spectrum, while it appears in absorption in all  other cases.
The absorption is deepest in the ULIRG spectrum and
shallowest in the Seyfert 1 spectrum. 
We illustrate the significant differences with example
spectra from the four classes 
(Figure \ref{fig:iras08572}).

We quantify the silicate feature, defining the ``feature strength''
\begin{equation}
S_{sil} = \ln {F_{obs}(\lambda) \over F_{cont}(\lambda)}
\end{equation}
in terms of the the observed ($F_{obs}$) and 
underlying continuum ($F_{cont}$)
flux evaluated at wavelength $\lambda$.
To determine the continuum well, we fit each
spectrum over the rest wavelength
intervals 5.0--7.0\um, 14.0--14.5\um, and 25.0--40.0\um.
We perform a linear fit  
to each of these three segments, 
measuring $\log F_\nu$ vs. $\log \lambda$,
then fit a spline to connect them.
We apply this approach directly to the
continuum-dominated  spectra, which include
most of the quasars.
In other cases, however, 
we modify the procedure to accomodate
spectral features, such as
high-ionization emission lines, 
polycyclic aromatic hydrocarbon (PAH) emission, and
ice absorption, 
excluding these regions
from the continuum fits.
For example, 
we shift the lower bound of the long-wavelength bandpass
to 26.5\um{} to avoid 
[\ion{O}{4}]$\lambda 25.9\um$ in many of the Seyfert galaxies. 
When PAH emission is strong,
the short-wavelength bandpass is reduced to 
a continuum measurement at 5.5\um,
and in ice-dominated spectra, we use continuum
measurements at 5.2 and 5.6\um.
(See Spoon et al. 2006b for detailed  \nocite{Spo06b}
examples of the continuum fits.)

We measure the strength $S_{sil}$ at the extremum 
of the continuum-subtracted spectrum.
Positive values of $S_{sil}$ indicate emission, while
negative values of $S_{sil}$ indicate absorption.
In absorption, this definition of feature strength 
corresponds to the apparent optical
depth, \textit{i.e.}, attenuation by a factor of $e^{-\tau_\lambda}$,  with
 $\tau_{sil,app}\ (= -S_{sil})$ conventionally evaluated at $\lambda = 9.7\um{}$
\citep[e.g.][]{Pie93,Gra94}.
The four classes show characteristic differences in strength, with
$S_{sil} = 0.17$, -0.18, -0.54, and -1.5 in the average spectra of the 
quasars, Seyfert 1 galaxies, Seyfert 2 galaxies, and ULIRGs,
respectively.
The most extreme example,
IRAS 08572+3915, has $S_{sil} = -4.0$  \citep{Spo06}.
While the ULIRGs exhibit a range of $S_{sil}$, including
a few cases of silicate emission where AGNs are identified optically,
the ULIRGs are generally deeply absorbed, with
$S_{sil} \le -1$ in 67\% 
and $S_{sil} \le -2$ in 30\% of this archival sample
 \citep{Hao06}. 
Moreover, all galaxies having $S_{sil} < -2$  are ULIRGs.

\section{Computation of the Spectral Energy Distribution\label{sec:computation}}
A purely absorbing
foreground screen cannot produce 
a deep silicate feature because 
the dust emission of an optically thick screen
is much greater than the transmitted external radiation. 
Denote by $B_\lambda$ the  Planck
function.
Dust emission
at wavelength $\lambda$ from a region with optical depth $\tau_\lambda$ is
$B_\lambda(T)(1 - e^{-\tau_\lambda})$ if the temperature $T$ is constant. 
Because the dust intensity is
$B_\lambda$ when the source is optically thick 
and $B_\lambda\tau_\lambda$ when
it is optically thin,  uniform temperature dust will never produce an
absorption feature under any circumstances. 
A temperature gradient is essential
for an absorption feature. 
Deep absorption requires dust geometry conducive to
large gradients. 

Consider a cloud illuminated from outside by a radiative source.
If  the cloud's dimensions are much smaller than the distance
to the source, then the heating flux is constant across the
cloud's volume.
Without radiative transfer effects, the dust temperature is uniform
throughout the cloud. 
In contrast, arranging the
same dust in a geometrically thick shell around the same
heating source produces a
large temperature gradient because of the spatial dilution of the flux with
radial distance. 
Radiative transfer
in optically thick dust affects
both geometries  similarly: the external radiation is absorbed within a short
distance from the illuminated face.
This 
creates a
large temperature differential close to the surface but only a 
modest gradient
in the deeper layers.
We can therefore expect the absorption feature to have only
limited depth in the case of externally illuminated clouds. 
{\em A deep feature
requires the radiation source to be deeply embedded in dust 
that is thick both
optically and geometrically}.

To verify this conclusion,  
we performed exact radiative transfer calculations
with 
the code DUSTY \citep{Ive99}.
The slab geometry serves as a proxy for an externally illuminated cloud, and a
spherical shell represents deeply embedding dust. 
We consider
a range of total optical depths, described in terms of
the optical depth at 5500\AA, $\tau_V$.
The SED of the underlying source spectrum is 
a broken power law, characteristic
of an AGN.  
In fact, as \citet{Row80} and \citet{Ive97} demonstrate
for similar problems, the resulting mid-IR spectral shape is 
insensitive to the heating spectrum.  
The mid-IR spectrum is ultimately a product of the dust,
and IR colors indicate its 
geometry \citep{Ive00}.

The model dust is that of the Galaxy, composed of 53\% silicate and
47\% graphite.  We use the absorption and scattering
cross sections of \citet{Dra03} and
the grain size distribution of  \citet{Mat77}.
Changing the silicate fraction does not alter any of the
general trends we report.
Crystalline silicates, however, produce sharper features in
the resulting SED, and we do not consider them further.

In both the geometrically thin slab and spherical shell 
cases, the
inner edge of the obscuring region is located where dust sublimates.
Assuming  $T_{sub} = 1500$ K, 
this inner radius $R_{in} \approx 1 L_{12}^{1/2}$ pc, where
$L_{12}$ is the luminosity of the central source in units of 
$10^{12} L_\sun$. 
While our concern here is the nuclei of galaxies, the results
are equally applicable to other situations, such as dust-embedded
stars.  The total optical depth completely describes the
slab.  For the spherical shell 
we consider variations of the  geometric thickness 
and the density distribution.

\section{Model Results\label{sec:results}}

\begin{figure}
\includegraphics[width=\hsize]{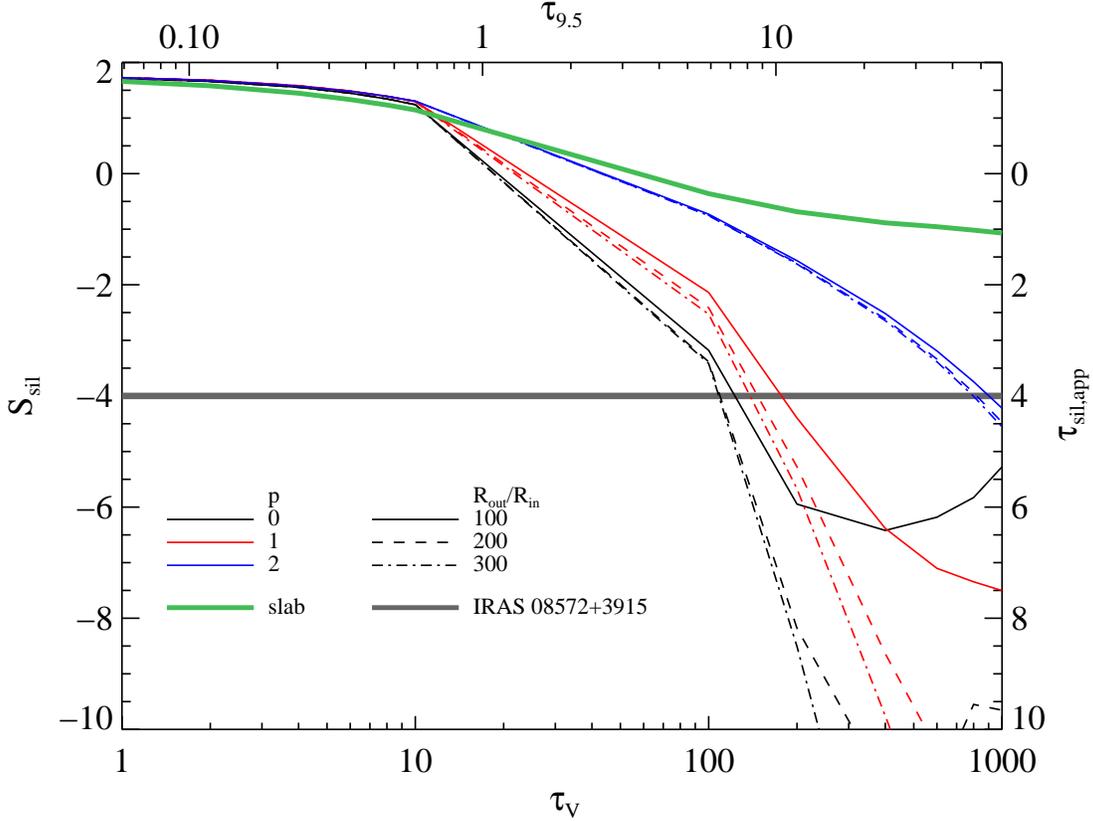}
\caption{\label{fig:strength}
Strength of the 10\um{} silicate feature (eq. 1) vs. optical depth (measured
at 5500\AA) for a slab and several spherical shell models.
The corresponding optical depth at 9.5\um{} is indicated along the
upper axis. Emission results in positive values of $S_{sil}$; 
absorption results in negative values.
 The apparent silicate optical depth, 
$\tau_{sil,app} \ ( = -S_{sil})$, 
when the feature appears in absorption
is indicated along the right axis.
The apparent optical depth is never large for the
 slab (\textit{thick green line}).
Thin spherical shells are similar to the slab result.
Several thick spherical shell models are plotted
(\textit{thin lines}), with the 
radial density profile ($\propto r^{-p}$)  and the 
shell thickness ($R_{out}/R_{in}$) indicated.
The  thick gray line shows 
$S_{sil}$ for IRAS 08572+3915.
Very large absolute values of $S_{sil}$, which
correspond to large apparent optical depth, are possible only
with the shell geometry.
}
\end{figure}

Figure \ref{fig:strength} summarizes the model results, showing
the 10\um{} feature strength, $S_{sil}$, vs. $\tau_V$
for the slab and several spherical shell models.
In the slab, the deepest absorption
approaches $S_{sil} = -1.1$ at $\tau_V =1000$. 
With the adopted extinction curve, 
$\tau_{9.5}/\tau_V =  0.06$, 
so the actual optical depth $\tau_{9.5} = 60$
while the apparent optical depth 
is only 1.1.
The depth of the feature does not
provide an accurate measurement of the total optical depth
along the line of sight, and the slab geometry
cannot produce the large observed apparent optical depths.  

The spherical shell results depend on 
the total shell thickness and the density profile, as
well as the total optical depth.
We considered 
 shell thickness, $R_{out}/R_{in}$, ranging from 1.5 to 1000.
Very thin shells are similar to the slab solutions, so
we plot only  representative thick shells, 
with outer radius $R_{out} = 100$--$300R_{in}$.
The radial density distribution is a power law, 
proportional to $r^{-p}$.
Thick spherical shells {\em can} produce extremely
deep silicate features, with the magnitude of
the feature strength increasing with decreasing $p$
and increasing $R_{out}$.
Even with the spherical geometry, the apparent optical depth
systematically underpredicts
 the actual line-of-sight optical depth. 

\begin{figure}
\includegraphics[width=\hsize]{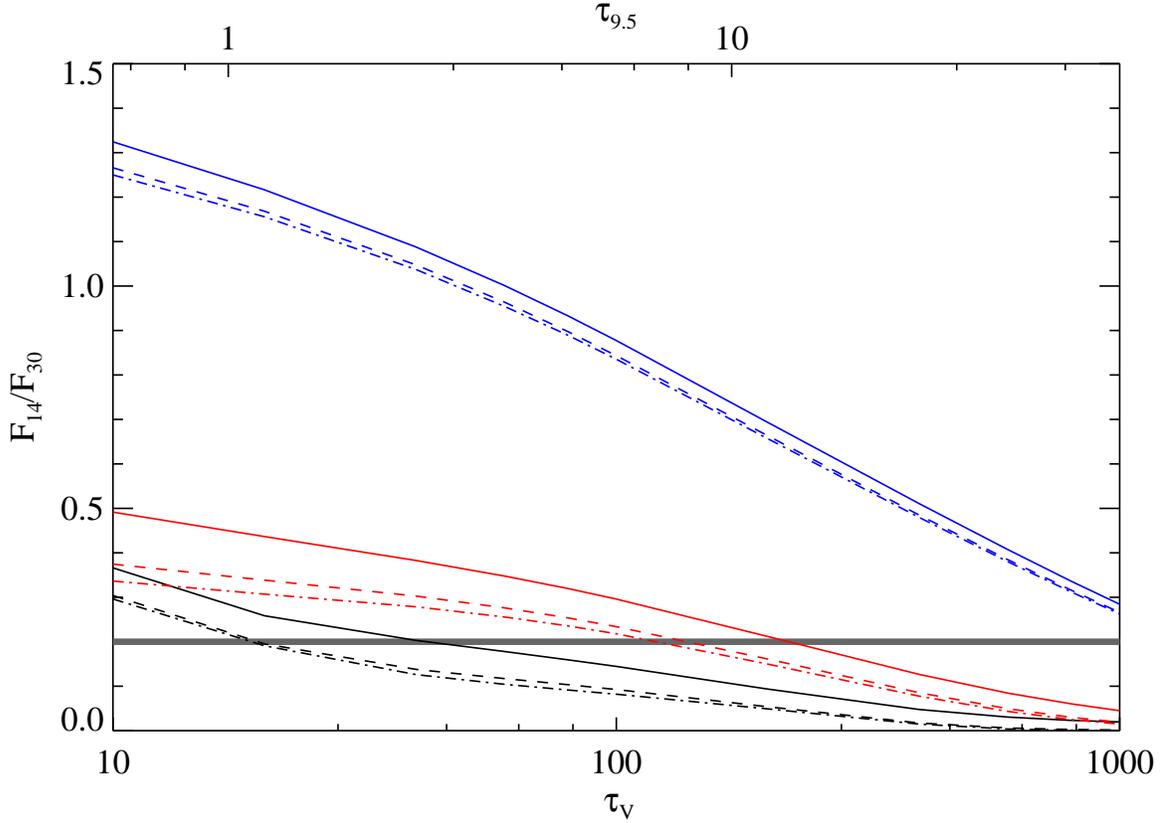}
\caption{\label{fig:specind}
Flux ratio between 14 and 30\um{} vs. optical depth.
Colors and line styles as in Fig. \ref{fig:strength}. 
The combination of $S_{sil}$ and this mid-IR color limits the  
acceptable models.
We observe $F_{14}/F_{30} = 0.20$ in IRAS 08572+3915,
which rules out  all $p=2$ (\textit{blue}) spectra.
Given the large strength,
the preferred solutions have $p= 1 $ (\textit{red}), with
$R_{out}/R_{in}= $ a few hundred, and $\tau_V =$ a few hundred.
}
\end{figure}

Several combinations of optical depth and shell geometry
can produce the large strength we observe in IRAS 08572+3915.  
A flux ratio helps to distinguish among them.
We compare 
$F_{\nu}$ 
at 14 and 30\um, where
the continuum dominates the observed flux.
As Figures \ref{fig:strength} and \ref{fig:specind}
together  illustrate, 
models with $p=1$,
moderate thickness 
($R_{out} \sim $ a few hundred $R_{in}$),
and moderately large optical depth ($\tau_V \sim$ a few hundred)
agree with 
both the observed  mid-IR color 
and the large magnitude $S_{sil}$ of IRAS 08572+3915.
None of the $p=0$ or 2 solutions 
fits both the feature depth and the 14/30\um{} flux ratio.  
However, the $p=0$ solutions at lower optical depth ($\tau_V\sim 30$)
are consistent with the average values of both  $S_{sil}$ and flux ratio
measured in the ULIRGs.

The observed flux densities separately 
indicate the intrinsic luminosity of the buried source. 
Comparing the  observations of IRAS 08572+3915 
with the simulated 
SED of the preferred solution, in which 
$\tau_{V} = 200$, $p =1$, and 
$R_{out} = 300R_{in}$,
the  14 and 30\um{} measurements
yield $L_{12} = 1.5$ and 1.1, respectively.
Both values are comparable to 
the luminosity derived from the IRAS fluxes,
$L_{12} = 1.4$, at a distance of 250 Mpc.
We also calculate the mass of the obscuring shell, 
adopting a standard gas-to-dust ratio $N_H = 2\times 10^{21} \tau_V$.
We find $M = 200 \tau_V L_{12} I M_\sun$,
where  $I$ depends
on the radial profile.  In terms of shell thickness, $Y= R_{out}/R_{in}$,
$I = Y^2/3$ for $p = 0$, $I=Y^2/(2 \ln Y)$ for $p =1$, and $I=Y$ for $p =2$.
Thus, in IRAS 08572+3915, 
$M \simeq 4\times 10^8 M_\sun$.

\section{Discussion and Conclusions\label{sec:concl}}
The strength of the absorption feature is a function of the
temperature contrast in the dusty material.
The extremely deep features we observe in many ULIRGs require 
a steep temperature gradient.
We can broadly account for the differences among
the spherical models in terms of the temperature
distribution of the dust.
In the limit of a thin shell ($R_{out}\to R_{in}$), 
the result approaches the slab solution. 
The geometric dilution of the heating is also reduced 
when the material is very concentrated (i.e., $p$ is large).
In general, the apparent optical depth increases
with  true optical  depth.
For a given radial density profile, 
the larger spheres typically exhibit deeper features 
because the temperature range over them is large. 
The exceptions to these general trends are a consequence
of the details of radiative transfer through the material,
and the results are sensitive to 
both temperature variation and optical depth. 
For example, 
any shell will become  heated well throughout
at sufficiently large optical depth,
reducing the temperature gradient and feature strength.
The onset of this effect depends on shell thickness  and $p$;
it is apparent in the $R_{out}/R_{in}$ = 100 solutions
for $p=0$, and to a lesser extent for $p=1$. 
We find the deepest absorption in
the $p=0$ or 1 models with moderate size.

We checked the effect of the chemical composition
of the dust, 
varying the silicate fraction from 20 to 80\%.
The general trends of $S_{sil}$
with geometry and optical depth remain the same 
in these additional simulations.
{\em Increasing the silicate abundance does not
produce a deeper feature};
only the
scaling of strength relative to total optical depth
 changes with
chemical composition 
because the ratio of $\tau_{9.5}$ to
total optical depth  changes.

The slab temperature profile is similar to that of any
externally heated cloud whose dimensions are much smaller
than its distance to the source.
The dark  and bright faces of the slab are distinct:
the silicate feature appears in absorption
only from the dark side, 
while viewing the bright
side always results in an emission feature, independent of the cloud's
total optical depth.
The spectrum of a clumpy medium mixes the contributions
of many individual clouds, 
each of which presents to the observer
either its dark or bright side 
\citep*{Nen02}.
Only a contrived cloud configuration can
produce a very deep absorption feature.
In a single slab,
the feature is deepest
when the temperature on the illuminated face is
around 200--400 K and the slab is viewed from its dark
side along the line of sight to the source. 
Unlike 
dust sublimation, there is no reason to single out
the distance corresponding to this particular temperature  
for the  location of all clouds.
Similarly, there is no reason for the observer
to be in a preferred
direction, one that views the dark sides exclusively.
In contrast, only the dark side of a  closed
shell can be observed.
However, once the shell is fragmented into clouds, 
it offers some direct views of portions of its
illuminated inner surface.
The emission from the bright faces fills
the absorption trough, 
reducing the depth of the  feature.

The true distribution of obscuring material is most
likely inhomogeneous, consisting of both dense clouds 
and a smooth intercloud medium, but the relative importance
of these two components differs between the ULIRGs and other
galaxies. 
The lack of deep absorption implies that the 
cloud-dominated medium is most relevant to
the AGNs.
Indeed, at  high spatial resolution, where the
AGN is isolated, 
the silicate absorption 
is never deep \citep{Roc06},
and complete models of obscuration by clouds can reproduce
the feature in detail \citep{Mas06}.
Individual Seyfert galaxies observed with \textit{Spitzer}
may show deep absorption when the larger (up to 20-kpc scale)
galactic contribution becomes comparable to that of the AGN.
Nonetheless, the absence of extreme absorption 
in the composite spectra of both Seyfert 1 and Seyfert 2 galaxies \citep{Hao06}
indicates that clouds dominate the obscuration of
their AGNs.

The deep silicate absorption that is 
characteristic of many ULIRGs
 requires a
central source embedded in
obscuring material 
that is thick
both 
optically and geometrically.
\citet{Ima03} noted 
similarly  that only 
centrally-concentrated 
buried  sources can produce 
deep 3\um{} dust absorption features.
While the medium may be inhomogeneous, the smooth
component of the distribution 
must dominate the 10\um{} spectral region.
Otherwise, emission from bright cloud faces would
fill in the absorption trough, reducing the magnitude
of $S_{sil}$.
Given the scaling of the shell size with luminosity, 
the observed deep absorption features therefore indicate 
obscuration on scales of a few 100 pc. 
AGNs may be present in these galaxies, but they must be
deeply buried.
The distinction of these ULIRGs is the continuous geometry of
the covering medium, not 
the optical depth along the line of sight or the
average covering fraction.
Indeed, 
 many Seyfert 2 galaxies
have large column densities along the lines of sight to
their nuclei, but they offer clear views of their
narrow line regions and some illuminated cloud faces.

The geometry of the obscuration determines the
emergent SED across the entire IR bandpass.
Our simulations demonstrate 
that 
the far-IR spectral shapes change significantly
with the geometric variations, despite having
the same heating source.
In general, the geometries that produce the deepest silicate
absorption also have the reddest far-IR spectra,
yet 
even modest optical depths ($\tau_V = 100$) can
have the peak of $F_\nu$ at wavelengths 
greater than 60\um. 
Clumpy and smooth distributions produce different
far-IR signatures \citep{Nen02}.
In a normal AGN, 
a clumpy medium (the torus)
obscures the central engine, and
the resulting 
far-IR flux ratios successfully
separate AGNs from  star-forming galaxies
\citep{deG87}.
However,
the nature of the heating source is erased by
dust reprocessing.  The only qualification 
is that the source provide UV photons to
heat the dust, which both starburst and AGN spectra do.
When an AGN is embedded in a thick,
smooth, dusty blanket, 
its peak flux density shifts to longer wavelengths.
Far-IR flux ratios 
cannot be used
to identify the powerful sources of radiation in 
the deeply-embedded ULIRGs.

Because of the extreme obscuration,
the emergent near-IR emission of the deeply embedded 
nuclear sources is weak.
The observed near-IR emission of ULIRGs 
is often stronger than
the deeply embedded models predict.
Even in IRAS 08572+3915, where the SED declines
steeply below 8\um{} (Fig. 1),
the near-IR emission exceeds the model results. 
Here 
the observed 3.7/14\um{} flux ratio is 0.05, but
the model embedded source 
does not account for more
than 5\% of the observed flux at $L$.
In these cases,  the component
responsible for the  observed near-IR emission must be different from
the one dominating at mid-IR wavelengths. 
The additional near-IR sources do not fill in the
deep silicate absorption troughs, so their spectra
must decline steeply above 5\um.  
The photospheric emission
of stars may therefore be the origin
of the observed near-IR light.
Because different sources dominate the observed
flux 
in different spectral regions,
a  near- to mid-IR flux ratio
is not generally 
a measurement of a single physical entity.

In summary, we have observed characteristic differences in 
the mid-IR silicate feature among four classes of galaxies.
The underlying physical distinctions of optical depth and
obscuring geometry account for these differences.
Optically thin regions produce silicate emission, which we
observe in the composite quasar spectrum.
Modest silicate absorption shows that clumpy obscuration
dominates in the Seyfert 1 and Seyfert 2 galaxies.
The extremely deep absorption in many ULIRGs, especially
that of IRAS 08572+3915, requires their  nuclear sources
to be deeply embedded in 
optically and geometrically thick material.

\acknowledgements
We thank D. Weedman for contributions to this work
and an anonymous referee for useful suggestions. 
NAL acknowledges work supported by the NSF 
under Grant No. 0237291.  ME acknowledges
support from NSF AST-0507421 and NASA NNG05GC38G. 

{\it Facility:} \facility{Spitzer (IRS)}

\end{document}